\documentclass[12pt]{article}
\usepackage{amsmath}
\usepackage[dvips]{graphics}
\DeclareFontFamily{OT1}{rsfs}{}
\DeclareFontShape{OT1}{rsfs}{m}{n}{ <-7> rsfs5 <7-10> rsfs7 <10->
rsfs10}{} \DeclareMathAlphabet{\mycal}{OT1}{rsfs}{m}{n}
\def\scri{{\mycal I}}%

\begin{document}
\newcommand{\bea}{\begin{eqnarray*}}
\newcommand{\eea}{\end{eqnarray*}}
\newcommand{\bean}{\begin{eqnarray}}
\newcommand{\eean}{\end{eqnarray}}
\newcommand{\eq}[1]{Eq. (\ref{#1})}
\newcommand{\meq}[1]{(\ref{#1})}
\newcommand{\eqn}[1]{(\ref{#1})}

\def\be{\begin{equation}}
\def\ee{\end{equation}}
\def\bea{\begin{eqnarray}}
\def\eea{\end{eqnarray}}
\def\nn{\nonumber}
\def\m{\mu}
\def\ga{\gamma}
\def\lan{\langle}
\def\ran{\rangle}
\def\Ga{\Gamma}
\def\thet{\theta}
\def\la{\lambda}
\def\Lam{\Lambda}
\def\ka{\chi}
\def\si{\sigma}
\def\al{\alpha}
\def\De{\Delta}
\def\ov{\over}

\newcommand{\tri}{\delta}
\newcommand{\grad}{\nabla}
\newcommand{\pa}{\partial}
\newcommand{\pf}[2]{\frac{\pa #1}{\pa #2}}
\newcommand{\bm}{\mathbf}
\newcommand{\de}{\delta}
\newcommand{\pp}{\frac{\partial}{\partial p}}
\newcommand{\pE}{\frac{\partial}{\partial E}}
\newcommand{\ppa}{\frac{\partial}{\partial p'}}
\newcommand{\pEa}{\frac{\partial}{\partial E'}}

\newcommand{\om}{\omega}
\newcommand{\omo}{\omega_0}
\newcommand{\ep}{\epsilon}
\newcommand{\nonu}{\nonumber}
\newcommand{\scrip}{\scri^{+}}
\newcommand{\hp}{{\cal H^+}}

\newcommand{\lxi}{{\cal L}_\xi}
\newcommand{\lt}{{\cal L}_t}
\newcommand{\lchi}{{\cal L}_\chi}
\newcommand{\psig}{{\partial\Sigma}}
\newcommand{\dbar}{\bar\delta}
\newcommand{\La}{\Lambda}
\newcommand{\sh}{S_{\cal H}}

\begin{center}
{\Large\bf The Energy of Bianchi Type I and II Universes in
Teleparallel Gravity}
\end{center}

\begin{center}
Lau Loi So\footnote{e-mail address: s0242010@cc.ncu.edu.tw},\ \
T. Vargas\footnote{e-mail address: teovar@ift.unesp.br}\\
$^{1}$Department of Physics, National Central University, Chung-Li
320, Taiwan.\\
$^{2}$ Instituto de Fisica Teorica, Universidade Estadual
Paulista, Rua Pamplona 145, 01405-900 Sao Paulo SP, Brazil.
\end{center}

\begin{abstract}
For certain models, the energy of the universe which includes the
energy of both the matter and the gravitational fields is obtained
by using the quasilocal energy-momentum in teleparallel gravity.
It is shown that in the case of the Bianchi type I and II
universes, not only the total energy but also the quasilocal
energy-momentum for any region vanishes independently of the three
dimensionless coupling constants of teleparallel gravity.

\end{abstract}

\section{Introduction}
The localization of energy and momentum of the gravitational field
is one of the oldest and most controversial problems of the
general theory of relativity~\cite{mtw}. After the energy-momentum
pseudotensor of Einstein~\cite{tra}, several other prescriptions
have been introduced, leading to a great variety of expressions
for the energy-momentum pseudotensor of the gravitational
field~\cite{lal}. These pseudotensors are not covariant objects
because they inherently depend on the reference frame, and thus
cannot provide a truly physical local gravitational
energy-momentum density. Consequently, the pseudotensor approach
has been largely questioned, although never abandoned. More
recently the idea of quasilocal (i.e. associated with a closed
2-surface) energy-momentum has become popular.  A large number of
definitions of quasilocal mass have been proposed~\cite{bryo}.

It has been shown recently by Chen and Nester~\cite{nest} that
every energy-momentum pseudotensor can be associated with a
particular Hamiltonian boundary term, which in turn determines a
quasilocal energy-momentum. In this sense, it has been said that
the Hamiltonian quasilocal energy-momentum rehabilitates the
pseudotensor approach, and dispels the doubts about the physical
meaning of these energy-momentum complexes.

However, Bergqvist~\cite{ber} studied several different
definitions of quasi-local masses for the Kerr and
Reissner-Nordstr\"om spacetimes and came to the conclusion that
not even two of these definitions gave the same result. On the
other hand, several authors studied various energy-mementum
complexes and obtained stimulating results. Virbhadra and
collaborators~\cite{virb} have demonstrated, with several
examples, that for many spacetimes (like the Kerr-Newman, Vaidya,
Einstein-Rosen, Bonnor-Vaidya and all Kerr-Schild class
spacetimes), different energy-momentum complexes give the same and
acceptable energy distribution for a given spacetime.

Despite this difficulty, there has been several attempts that
calculated the total energy of the expanding universe.
Investigations~\cite{alb} and~\cite{tryo} proposed that our
universe may have arisen as a quantum fluctuation of the vacuum,
and mentioned that any conservation law of physics need not have
been violated at the time of its creation. Tryon  proposed that
our universe must have a zero net value for all conserved
quantities and presented some arguments, using a Newtonian order
of magnitude estimate, favoring the fact that the net energy of
our universe may be indeed zero. Their model predicts a universe
which is homogeneous, isotropic and closed, and consists equally
of matter and anti-matter. The subject of the total energy of the
expanding universe was re-opened by~\cite{coop,ros,other}, using
the definitions of the energy-momentum in general relativity. In
one of these the Einstein energy-momentum pseudotensor was used to
represent the gravitational energy~\cite{ros}, which led to the
result that the total energy of a closed Friedman-Robertson-Walker
(FRW) universe is zero. In another, the symmetric pseudotensor of
Landau-Lifshitz was used~\cite{other}. In~\cite{base}
and~\cite{raxu}, the total energy of the anisotropic Bianchi
models have been calculated using different pseudotensors, leading
to similar results. Recently~\cite{facoo}, it has been shown that
the open, or critically open FRW universes, as well as Bianchi
models evolving into de Sitter spacetimes also have zero total
energy. Finally, a similar calculation for the closed FRW universe
using the Einstein and Laudau-Lifshitz complexes in teleparallel
gravity also led to the same conclusion~\cite{var}.  There is a
well known argument that the energy of a closed universe should
vanish. But the argument does not apply to open universes.

By working in the context of teleparallel gravity, it will be
shown that not only the total but also the quasilocal energy for
any region of the Bianchi type I and II universes vanishes,
independently of the three dimensionless coupling constants of
teleparallel gravity. It should be remarked that this result is
consistent with the results of Banerjee and Sen~\cite{base}, Xulu,
and Radinschi~\cite{raxu} calculated in the framework of general
relativity. We will proceed according to the following scheme. In
section~2, we review the main features of teleparallel gravity and
the expression for the quasilocal energy-momentum in teleparallel
gravity. In section~3, we find the tetrad field, the non-zero
components of the Weitzenb\"ock connection, the torsion tensor,
and then the quasilocal energy and momentum of the Bianchi type I
and II universes are calculated. Finally, in section~4, we present
a discussion of the result obtained.

\section{Teleparallel Gravity}

In teleparallel gravity, spacetime is represented by the Weitzenb\"{o}ck
manifold $W^{4}$ of distant parallelism~\cite{we}. This gravitational theory naturally
arises within the gauge approach based on the group of the spacetime
translations. Denoting the translational gauge potential by $A^{a}{}_{\mu}$,
the gauge covariant derivative for a scalar field $\Phi(x^{\mu})$ reads~\cite{pe}
\begin{equation}
D_{\mu}\Phi=h^{a}{}_{\mu}\partial_{a}\Phi,
\label{drc}
\end{equation}
where
\begin{equation}
h^{a}{}_{\mu} = \partial_{\mu}x^{a} + A^{a}{}_{\mu}
\label{tetra}
\end{equation}
is the tetrad field, which satisfies the orthogonality condition
\begin{equation}
h^{a}{}_{\mu}\,h_{a}{}^{\nu}=\delta^{\nu}_{\mu} .
\label{ort}
\end{equation}
The nontrivial tetrad field induces a teleparallel structure on
spacetime which is directly related to the presence of the
gravitational field; the Riemannian metric arises as
\begin{equation}
g_{\mu \nu} = \eta_{a b} \; h^a{}_\mu \; h^b{}_\nu \; .
\label{met}
\end{equation}
In this theory, the fundamental field is a nontrivial tetrad, which gives
rise to the metric as a by-product. The parallel transport of the tetrad
$h^{a}{}_{\mu}$ between two neighbouring points is encoded in the covariant
derivative
\begin{equation}
\nabla_{\nu}h^{a}{}_{\mu}=\partial_{\nu}h^{a}{}_{\mu}
-\Gamma ^{\alpha}{}_{\mu \nu}h^{a}{}_{\alpha},
\label{nt}
\end{equation}
where $\Gamma^{\alpha}{}_{\mu \nu}$ is the  Weitzenb\"ock connection, a
connection presenting torsion, but no curvature. Imposing the condition
that the tetrad be parallel transported in the Weitzenb\"ock space-time, we
obtain
\begin{equation}
\Gamma ^{\alpha}{}_{\mu \nu}=h_{a}{}^{\alpha}\partial_{\nu}h^{a}{}_{\mu},
\end{equation}
which gives the explicit form of the Weitzen\-b\"ock connection in
terms of the tetrad;
\begin{equation}
T^{\rho}{}_{\mu \nu}=\Gamma ^{\rho}{}_{\nu \mu}-\Gamma ^{\rho}{}_{\mu \nu}
\end{equation}
is the torsion of the Weitzenb\"ock connection. As we already remarked,
the curvature of the Weitzenb\"ock connection vanishes identically as a
consequence of absolute parallelism, or teleparallelism~\cite{xa}.

The action of teleparallel gravity in the presence of matter is given by
\begin{equation}
S = \frac{1}{16 \pi G} \int d^{4}x \, h\,S^{\la \tau \nu} \;
T_{\la \tau \nu}+ \int d^{4}x \, h\,{\mathcal L_{M}} \label{lag},
\end{equation}
where $h = \det(h^a{}_\mu)$, $\mathcal L_{M}$ is the Lagrangian of
the matter field, and $S^{\la \tau \nu}$ is the tensor
\begin{equation}
S^{\la \tau \nu} = c_1T^{\la \tau \nu}+\frac{c_2}{2}\left(T^{\tau \la \nu} -
T^{\nu \la \tau}\right) +\frac{c_3}{2}\left(g^{\la \nu} \;
T^{\si \tau}{}_{\si} - g^{\tau \la} \;T^{\si \nu}{}_{\si} \right),\label{2.32a}
\end{equation}
with $c_1$, $c_2$ and $c_3$ being the three dimensionless coupling
constants of teleparallel gravity~\cite{xa}. For the specific
choice
\begin{equation}
c_1=\frac{1}{4}, \quad c_2=\frac{1}{2}, \quad c_3=-1,
\label{par}
\end{equation}
teleparallel gravity yields the so called teleparallel equivalent
of general relativity~\cite{maluf}.

By performing the variation in (\ref{lag}) with respect to
$h^{a}{}_{\mu}$, we get the teleparallel field equations,
\begin{equation}
\partial_\sigma(h S_{\lambda}{}^{\tau \sigma}) - 4 \pi G\;
(ht^{\tau}{}_{\la}) = 4 \pi G\;h\; T^{\tau}{}_{\la},     \label{eqc}
\end{equation}
where
\begin{equation}
t^{\tau}{}_{\la} = \frac{1}{4 \pi G}h\Gamma^{\nu}{}_{\sigma \lambda}
S_{\nu}{}^{\tau \sigma} - \delta^{\tau}{}_{\la}\mathcal L_{G}
\end{equation}
is the energy-momentum pseudotensor of the gravitational field~\cite{andrad}.
Rewriting the teleparallel field equations in the form
\begin{equation}
h(t^{\tau}{}_{\la} + T^{\tau}{}_{\la}) = \frac{1}{4 \pi G}
\partial_\sigma(h S_{\lambda}{}^{\tau \sigma}),     \label{pse}
\end{equation}
as a consequence of the antisymmetry of $S_{\lambda}{}^{\tau \sigma}$ in
the last two indices, we obtain immediately the conservation law
\begin{equation}
\partial_{\tau}[h (t^{\tau}{}_{\la}+T^{\tau}{}_{\la})] = 0.
\end{equation}
On the other hand, in the Hamiltonian formalism, the Hamiltonian
for a finite region
\begin{equation}
H(N)=\int_{\Sigma}N^{\mu}\mathcal{H}_{\mu}+\oint_{\partial \Sigma}\mathcal{B}(N)
\end{equation}
generates the spacetime displacement of a finite spacelike
hypersurface $\Sigma$ along a vector field $N^{\mu}$. Noether's
theorems guarantee that $\mathcal{H}_{\mu}$ is proportional to the
field equations. Consequently, the value depends only on the
boundary $\mathcal{B}$, which gives the quasilocal
energy-momentum~\cite{nest}.  The boundary term $\mathcal{B}(N)$
in teleparallel gravity~\cite{nest2,andrad,maluf} is given by
\begin{equation}
\mathcal{B}(N)=\frac{1}{8\pi
G}N^{\mu}hS_{\mu}{}^{\alpha\beta}(d\si)_{\alpha\beta}.
\end{equation}
Thus, the quasilocal energy-momentum in teleparallel gravity is
\begin{equation}
P_{\mu}=\frac{1}{8\pi G}\oint_{\partial
\Sigma}hS_{\mu}{}^{\alpha\beta}(d\si)_ {\alpha\beta}.
\label{quasi}
\end{equation}
From this equation, we see that the energy and other quasilocal
quantities are given by the integral of the $\mathcal{B}(N)$ over
the 2-surface boundary $\partial \Sigma$ of the 3-hypersurface
$\Sigma$.

\section{Calculation of the total energy}

In this section we will calculate the total energy of the Bianchi
type I and II space-times using the quasilocal energy-momentum in
teleparallel gravity given by equation (\ref{quasi}).
\begin{itemize}
\item
{\it \large{Bianchi type I space-time}}

The homogeneous and anisotropic Bianchi type I spacetimes are
expressed by the line element~\cite{base}
\begin{equation}
ds^{2}=dt^{2}-e^{2l}dx^{2}-e^{2m}dy^{2}-e^{2n}dz^{2},
\end{equation}
where l, m, n are functions of time t only. Using relation
(\ref{met}), we obtain the tetrad components and its inverse
\begin{eqnarray} h^{a}{}_{\mu}=
\left(\begin{array}{cccc}
1&0&0&0\\
0&e^{l}&0&0\\
0&0&e^{m}&0\\
0&0&0&e^{n}
\end{array}\right),
\ h_{a}^{\ \mu}= \left(\begin{array}{cccc}
1&0&0&0\\
0&e^{-l}&0&0\\
0&0&e^{-m}&0\\
0&0&0&e^{-n},
\end{array}\right).\nonumber
\end{eqnarray}
The corresponding non-vanishing Weitzenbock connection terms are:
\begin{equation} \Gamma^{1}_{\ 10}=\dot{l}, \quad\Gamma^{2}_{\ 20}=\dot{m} \quad and
\quad \Gamma^{3}{}_{30}=\dot{n},\nonumber
\end{equation}
where dot denotes the derivative with respect to time t.  Hence
the non-vanishing torsion components are:
\begin{eqnarray}
T^{1}_{\ \ 01}&=&-T^{1}_{\ \ 10}\ =\ \dot{l}\ ,\nonumber \\
T^{2}_{\ \ 02}&=&-T^{2}_{\ \ 20}\ =\ \dot{m}\ \nonumber,\\
T^{3}_{\ \ 03}&=&-T^{3}_{\ \ 30}\ =\ \dot{n}\nonumber.
\end{eqnarray}
Then the non-zero components of the tensor $S_{\nu}^{\
\sigma\tau}$ read:
\begin{eqnarray}
S_{1}^{\ 01}&=&-S_{1}^{\ 10}\ =\
\left(c_{1}+\frac{c_{2}}{2}\right)\dot{l} +
\frac{c_{3}}{2}\left(\dot{l}+\dot{m}+\dot{n}\right)\,,\nonumber\\
S_{2}^{\ 02}&=&-S_{2}^{\ 20}\ =\
\left(c_{1}+\frac{c_{2}}{2}\right)\dot{m} +
\frac{c_{3}}{2}\left(\dot{l}+\dot{m}+\dot{n}\right)\,,\nonumber\\
S_{3}^{\ 03}&=&-S_{3}^{\ 30}\ =\
\left(c_{1}+\frac{c_{2}}{2}\right)\dot{n} +
\frac{c_{3}}{2}\left(\dot{l}+\dot{m}+\dot{n}\right).
\label{tensor.S}
\end{eqnarray}
Note that there is an interesting identity:\
$S_{\alpha\beta\gamma}+S_{\beta\gamma\alpha}+S_{\gamma\alpha\beta}=0$.
Using Eq. (\ref{quasi}), the quasilocal energy within any region
$\Sigma$ is
\begin{equation}
P_{0}=\frac{1}{8\pi G}\oint_{\partial
\Sigma}hS_{0}{}^{\alpha\beta}(d\si)_ {\alpha\beta}=0,
\end{equation}
since $S_{0}{}^{\alpha\beta}=0$ for all $\alpha,\ \beta$. Our
calculation is consistent with the Radinschi and Xulu
results~\cite{raxu}. From Eq. (\ref{tensor.S}) it is also readily
follow that
\begin{equation}
P_{i}=\frac{1}{4\pi G}\oint_{\partial\Sigma}
hS_{i}{}^{0j}(d\sigma)_{0j}=0.
\end{equation}
\item
{\it \large{Bianchi type II space-time}}

The Bianchi type II line element ~\cite{metricII} is
\begin{equation}
ds^{2}=dt^{2}-f^{2}dx^{2}-g^{2}dy^{2}+2xg^{2}dydz-(x^{2}g^{2}+f^{2})dz^{2},
\end{equation}
where $f(t)$ and $g(t)$ depend only on time.

Using again the relation (\ref{met}), we obtain the tetrad components and its inverse
\begin{eqnarray} h^{a}_{\ \mu}=
\left(\begin{array}{cccc}
1&0&0&0\\
0&f&0&0\\
0&0&g&-xg\\
0&0&0&f
\end{array}\right),
\ h_{a}^{\ \mu}= \left(\begin{array}{cccc}
1&0&0&0\\
0&f^{-1}&0&0\\
0&0&g^{-1}&0\\
0&0&xf^{-1}&f^{-1}
\end{array}\right).\nonumber
\end{eqnarray}
The non-vanishing connections are:
\begin{eqnarray}
\Gamma^{1}{}_{10}&=&\Gamma^{3}{}_{30}\ =\ \frac{\dot{f}}{f}\,,\,
\ \ \Gamma^{2}{}_{20}=\frac{\dot{g}}{g}\,,\nonumber\\
\Gamma^{2}{}_{30}&=&x\left(\frac{\dot{f}}{f}-\frac{\dot{g}}{g}\right)\,,
\ \Gamma^{2}{}_{31}=-1\,.\nonumber
\end{eqnarray}
Modulo anti-symmetry the non-vanishing torsion components are:
\begin{eqnarray}
T^{1}{}_{01}&=&T^{3}_{\ 03}\ =\ \frac{\dot{f}}{f}\,,
\quad T^{2}{}_{02}=\frac{\dot{g}}{g}\,,\nonumber\\
T^{2}{}_{03}&=&x\left(\frac{\dot{f}}{f}-\frac{\dot{g}}{g}\right)\,,\,
\ T^{2}{}_{13}=-1\,.\nonumber
\end{eqnarray}
Modulo anti-symmetry the non-vanishing components of S are:
\begin{eqnarray}
S_{1}{}^{01}&=&S_{3}{}^{03}=\left(c_1+\frac{c_2}{2}
+c_3\right)\frac{\dot{f}}{f}+\frac{c_3}{2}\frac{\dot{g}}{g}\,,\nonumber\\
S_{1}{}^{23}&=&\frac{c_2}{2} \frac{1}{f^{2}}\,,
\quad \quad S_{2}{}^{12}=-S_{3}{}^{13}=c_{1}\frac{xg^{2}}{f^{4}}\,, \nonumber\\
S_{2}{}^{02}&=&\left(c_1+\frac{c_2}{2}
+\frac{c_3}{2}\right)\frac{\dot{g}}{g}+c_3\frac{\dot{f}}{f}\,,
\quad S_{2}{}^{13}=c_{1}\frac{g^2}{f^4}\,,\nonumber\\
S_{3}{}^{02}&=&\left(c_1+\frac{c_2}{2}\right)\left(\frac{\dot{f}}{f}
-\frac{\dot{g}}{g}\right)x \,, \quad
S_{3}{}^{12}=-c_{1}\frac{x^{2}g^{2}}{f^{4}}+\frac{c_2}{2}\frac{1}{f^{2}}\
. \label{tensor.S2}
\end{eqnarray}
Therefore the quasilocal energy within any region is
\begin{equation}
P_{0}=\frac{1}{8\pi G}\oint_{\partial
\Sigma}hS_{0}{}^{\alpha\beta}(d\si)_{\alpha\beta}=0,
\end{equation}
since $S_{0}{}^{\alpha\beta}=0$ for all $\alpha$ and $\beta$, a
result consistent with that of Banerjee and Sen~\cite{base}. Also,
using Eq. (\ref{tensor.S2}), a short calculation gives the
qusilocal momentum
\begin{equation}
P_{i}=\frac{1}{8\pi G}\oint_{\partial\Sigma}
hS_{i}{}^{\alpha\beta}(d\sigma)_{\alpha\beta}=0.
\end{equation}
\end{itemize}
\section{Final Remarks}

In order to compute the gravitational energy, we have considered
the teleparallel quasilocal energy-momentum. Working in the
context of teleparallel gravity, we have calculated the quasilocal
energy-momentum of the Bianchi type I and II universes. Our basic
result is that the quasilocal energy and momentum, which includes
the energy of both the matter and the gravitational fields
vanishes for all regions, not just for the total space.  It is
also independent of the three teleparallel dimensionless coupling
constants, which means that it is valid not only in the
telaparallel equivalent of general relativity, but also in any
teleparallel model. According to these calculations, the total
energy vanishes everywhere because the energy contributions from
the matter and gravitational fields inside an arbitrary
two-surface boundary $\partial \Sigma$ of the 3-hypersurface
$\Sigma$ cancel each other.

\section*{Acknowledgments}
The authors would like to thank Prof. J M Nester for fruitful discussions, and for a
crititcal reading of the manuscript. TV would like also to thank FAPESP for
financial support.

\end{document}